\documentclass[conference]{IEEEtran}
\usepackage{cite,graphicx,amsmath,amssymb}
\usepackage{subfigure}
\usepackage{fancyhdr}
\usepackage{mdwmath}
\usepackage{mdwtab}
\usepackage{balance}
\usepackage{xcolor}
\usepackage{bm}
\usepackage{amsmath}

\usepackage{algorithm}
\usepackage{algorithmic}
\usepackage{multirow}
\usepackage{flafter}
\usepackage{amssymb}
\usepackage{amsmath}
\usepackage{cite}
\usepackage{url}
\usepackage{xcolor}
\usepackage{cite,graphicx,amsmath,amssymb}
\usepackage{subfigure}
\usepackage{caption}

\usepackage{fancyhdr}
\usepackage{mdwmath}
\usepackage{mdwtab}
\usepackage{caption}
\usepackage{amsthm}
\usepackage{comment}

\usepackage[
top    = 1.70cm,
bottom = 1.05in,
left   = 0.63 in,
right  = 0.63 in]{geometry}

\def\BibTeX{{\rm B\kern-.05em{\sc i\kern-.025em b}\kern-.08em T\kern-.1667em\lower.7ex\hbox{E}\kern-.125emX}}
\newtheorem{lemma}{Lemma}

\newtheorem{remark}{Remark}
\newtheorem{theorem}{Theorem}

\newtheorem{corollary}{Corollary}

\begin{document}

\title{Reconfigurable Intelligent Surface-assisted Networks: Phase Alignment Categories}
\author{
\IEEEauthorblockN{  Jiaqi~Xu\IEEEauthorrefmark{1}, and Yuanwei~Liu\IEEEauthorrefmark{1}} \IEEEauthorblockA{\\
\IEEEauthorrefmark{1} Queen Mary University of London, London, UK\\
\{jiaqi.xu, yuanwei.liu\}@qmul.ac.uk\\
}}
\maketitle
\begin{abstract}
The reconfigurable intelligent surface (RIS) is one of the promising technology contributing to the next generation smart radio environment. The application scenarios include massive connectivity support, signal enhancement, and security protection. One crucial difficulty of analyzing the RIS-assisted networks is that the channel performance is sensitive to the change of user receiving direction. This paper tackles the problem by categorizing the RIS illuminated space into four categories: perfect alignment, coherent alignment, random alignment, and destructive alignment. These four categories cover all the possible phase alignment conditions that a user could experience within the overall $2$ pi solid angle of RIS-illuminated space. We perform analysis for each of these categories, deriving analytical expressions for the outage probability and diversity order. Simulation results are presented to confirm the effectiveness of the proposed analytical results.

\end{abstract}

\begin{IEEEkeywords}
Channel model, diversity order, non-orthogonal multiple access, phase errors, reconfigurable intelligent surface.
\end{IEEEkeywords}

\section{Introduction}

In the RIS-assisted wireless communication system, the RIS can perform signal enhancing (at a target angle), signal broadcasting, and signal cancelling\cite{liu2020reconfigurable}. These different working conditions are the results of different alignments of the phase shifters, and they are subject to a specific RIS-user pair. With sufficient channel information of the receivers, the RIS can assist in achieving massive connectivity, security protection, and more emerging application demands for future generation wireless networks.

Existing research contributions focused on the study of the RIS-assisted channel for the optimized user. The effect of RIS quantization levels on the channel distribution and diversity gain has been analyzed. In \cite{xu2020reconfigurable}, Xu~\textit{et al.} claimed that full diversity order can be achieved by RIS with a quantization level of $L=3$. In \cite{wang2020study}, Wang~\textit{et al.} showed the diversity order of the one-bit discrete phase shift RIS system is $(M+3)/2$, where $M$ is the number of elements of the RIS. In recent performance analysis works, it was pointed out that using the central limit theorem (CLT) causes a systematic error. In \cite{wang2020chernoff}, Wang~\textit{et al.} avoided the CLT by using Chernoff inequality and saddle-point approximation. In \cite{ding2020impact}, Ding~\textit{et al.} presented upper bounds of the outage probability for RIS with continuous phase shift. 

The main novelty that distinguishes this performance analysis work from other related research is that we study the RIS-assisted channel in all possible user directions. According to the law of energy conservation, the RIS redistributes the radiation power within different angles in the $2$-pi illuminated space. At positions other than that of the targeted user, the channels produce weaker averaged received powers. In fading environments, these channels also exhibit different distributions compared with the channel in the target direction.

This paper aims to analyze the channel outage probability for the four proposed phase alignment categories: perfect alignment, coherent alignment, random alignment, and destructive alignment. Table~\ref{table1} summaries the expected magnitude, variance, and diversity order (assuming Rayleigh or Rician faded diversity branches) of the RIS-assisted channel in different categories. It is one of the main takeaways of the paper. The structure of the paper is as follows: Firstly, we proposed the four categories and their mathematical descriptions. Then, we derive analytical expressions for the outage probability of the proposed phase alignment categories, considering both continuous phase shift RIS and discrete phase shift RIS scenarios. We utilize the Laplace transform and its convolution theorem to assist our analysis and demonstrate the achievable diversity order for each category. Finally, we perform Monte Carlo-based simulations to verify the effectiveness of the proposed analytical results.

\begin{table*}
\centering
\begin{tabular}{|c|c|c|c|c|}
\hline
Working conditions & \multicolumn{2}{c|}{Enhancing} & Broadcasting & Cancelling
\\ \hline
Phase alignment       & (a) Perfect             & (b) Coherent        & (c) Random            &(d) Destructive  \\ \hline
$\mathbb{E}[|H|]$       & $M\bar{h}$              &  $\sqrt{\pi/2}\beta L_{1/2}(-\alpha^2/(2\beta^2))$                  &$\sqrt{M\pi\bar{h^2}}/2$ & $0$                       \\ \hline
$Var[|H|]$              & $M(\bar{h^2}-(\bar{h})^2)$                       &$\alpha^2+2\beta^2-(\mathbb{E}[H])^2$                & $M\bar{h^2}(4-\pi)/4$                      & $M(\bar{h^2}-(\bar{h})^2)$ \\ \hline
Diversity order & \multicolumn{1}{c|}{M} & Less or close to M & \multicolumn{1}{c|}{1} & $0$ \\ \hline
\end{tabular}
\caption{Comparing different phase alignment categories,}
where $\bar{h}=\mathbb{E}[h_m]$, $\bar{h^2}=\mathbb{E}[h^2_m]$, all $h_m$ are independent and identically distributed, $\alpha = M\bar{h}sinc(\pi/(2L))$, $\beta^2 = M\bar{h^2}[1-sinc(\pi/L)]/2$, and $L_{1/2}(x)$ denoting the Laguerre polynomial.
\begin{proof}
See Appendix~\ref{ap_A} for the proof of expectation values and variances. Derivations for the diversity orders are presented in Section~\ref{asymp}.
\end{proof}
\label{table1}
\end{table*}

\section{Mathematical Descriptions of the Four Phase Alignment Categories in Fading Environment}\label{math_4}
In the problems arising in the smart (programmable) wireless communication environment, the resultant field is formed by a superposition of a number of elementary waves:
\begin{equation}\label{superposition}
\tilde{H} = |H|e^{j\phi_H}=\sum_{m=1}^M|h_m|e^{j\theta_m},
\end{equation}
where $|h_m|$ and $\theta_m$ are the amplitude and the phase of the $m$-th diversity branch. In the context of analyzing RIS performance, it is natural to associate index $m$ in \eqref{superposition} to each element on the RIS. Consider the presence of small-scale fading, both $|h_m|$ and $\theta_m$ may be random. In most practical cases, determine the exact distribution of $|H|$ is difficult. However, assuming specific distribution for $|h_m|$ and $\theta_m$, we are able to draw useful insights into the characteristics of the distribution of $H$, such as its mean value, variance, and asymptotic behaviour.

\begin{figure*}[ht!]
    \begin{center}
        \includegraphics[scale=0.7]{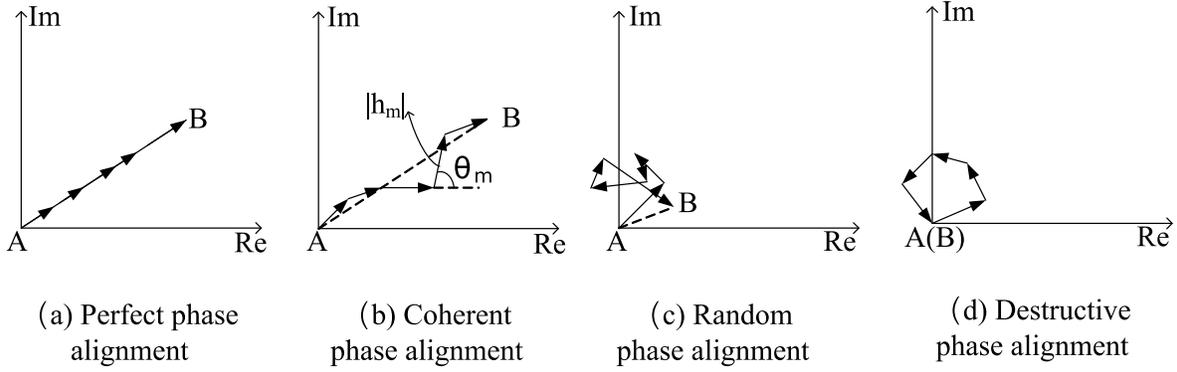}
        \caption{Phasor diagram of the four phase alignment categories.}
        \label{three}
    \end{center}
\end{figure*}

In the following analysis, we investigate the distribution of $H$ by choosing $|h_m|$ to follow Rayleigh fading or Rican fading. Different phase distribution of $\theta_m$ can be classified into four categories, as in Fig.~\ref{three}.
\subsubsection{Perfect phase alignment}
For perfect phase alignment, we have $\theta_m=\theta_0, \forall m\in [1,M]$. It indicates that $\theta_m$ is a fixed value without any distribution. Thus, the magnitude of $H$, which is the length of segment $AB$ in Fig.~\ref{three}, can be simplified as $|H|=\sum_{m=1}^M|h_m|$.

\subsubsection{Coherent phase alignment}
For coherent phase alignment, we have $\theta_m = \theta_0+\Delta\theta_m$. Various reasons are causing the phase error $\Delta\theta_m$, such as discrete phase shifts of the RIS, or imperfect CSI at the RIS so that the RIS is unable to perform perfect phase alignment. In different scenarios, $\Delta\theta_m$ distributes differently. However, the case where $\Delta\theta_m$ is uniformly distributed within $(-\pi/L,+\pi/L)$ is of special interest. The quantity of $L$ is often referred to as the quantization level in the literature. 

\subsubsection{Random and destructive phase alignment:}
For random phase alignment, $\theta_m$ has a uniform distribution within $(-\pi,\pi)$. For destructive phase alignment, the distribution of $\theta_m$ ensures that the magnitude of $H$ is zero.

\addtolength{\topmargin}{0.02in}
\section{Performance Analysis for the RIS-assisted Networks}\label{asymp}
\subsection{Uniform 1-D Phase Scanning}

\begin{figure}[ht!]
    \begin{center}
        \includegraphics[scale=0.3]{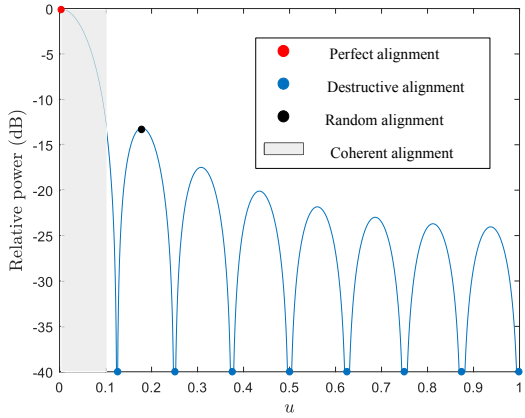}
        \caption{Illustration of phase alignment categories in the radiation pattern of 1-D phase scanning RIS.}
        \label{user_pers}
    \end{center}
\end{figure}
In the application scenarios where a targeted user is served by the RIS under line-of-sight links, the optimal phase configuration is the uniform 1-D phase scanning pattern. This configuration is suitable for serving users at a distinct direction ($\theta_0$) and it is achieved by configure the reflection coefficient of the $m$-th element as:
\begin{align}\label{c_scan}
r_m = exp(2\pi ju_0\frac{d_x}{\lambda} m),
\end{align}
where $u_0 = \sin\theta_0$ is the target direction. This phase RIS configuration dose not require amplitude control, since from \eqref{c_scan}, we have $|r_m|=1,\  \forall m \in [1,M]$. The radiation pattern, normalized to its peak value, can be expressed as~\cite{mailloux2017phased}:
\begin{equation}\label{pattern_0}
F(u)=\frac{\sin[M\pi d_x(u-u_0)/\lambda]}{M\sin[\pi d_x(u-u_0)/\lambda]}.
\end{equation}
Fig.~\ref{user_pers} illustrates four phase alignment categories in the radiation pattern of a 16-element 1-D phase scanning RIS. For the continuous phase shift RIS, perfect phase alignment can be achieved at the target direction, which corresponds to $u=0$ in the figure. However, for discrete phase shift RIS or RIS with imperfect channel state information (CSI), perfect phase alignment cannot be achieved. Table~\ref{table2} presents the corresponding phase alignment categories for different CSI conditions and RIS settings.

\begin{table}[!h]
\centering
\begin{tabular}{|c|c|c|}
\hline
RIS phase shift & CSI     & Phase alignment at target direction \\ \hline
Continuous      & Perfect & Perfect alignment                   \\ \hline
Continuous      & Partial & Coherent alignment                  \\ \hline
Continuous      & None    & Random alignment                    \\ \hline
Discrete        & Perfect & Coherent alignment                  \\ \hline
Discrete        & Partial & Coherent alignment                  \\ \hline
Discrete        & None    & Random alignment                    \\ \hline
\end{tabular}
\caption{Phase alignment categories of different RIS phase shift and CSI conditions, for 1-D phase scanning RIS configuration.}
\label{table2}
\end{table}

\subsection{Definition of Outage Probability and Diversity Order}
For the application scenarios in fading environment, the signal-to-noise ratio (SNR) of the combined received signal is a real-valued random variable. The outage probability is the CDF of the combined SNR:
\begin{equation}\label{out_define}
P_{out}(\gamma_0)=\int_{x=0}^{\gamma_0}P_{\gamma_\Sigma}(x)dx,
\end{equation}
where $\gamma_0$ is the targeted SNR threshold and $P_{\gamma_\Sigma}(x)$ is the PDF of the combined received SNR ($\gamma_\Sigma$). In our analysis, it is more convenient to rewrite the outage probability in terms of the PDF of the overall channel amplitude:
\begin{equation}\label{out_1}
P_{out}(\gamma_0)=\int_{0}^{x^{*}}P_{|H|}(x)dx,
\end{equation}
where $x^*=\sqrt{\gamma_0/\gamma_t}$, $\gamma_t$ is average the SNR per branch, and $P_{|H|}$ is the PDF of the overall channel amplitude ($H$).

The diversity order of the system can be defined through outage probability as:
\begin{equation}\label{order_define}
d=-\lim_{\gamma_t \to \infty}\frac{\log P_{out}(\gamma_0)}{\log \gamma_t}.
\end{equation}

\subsection{Methodology}
In the analysis of the RIS-assisted wireless channel, the overall channel amplitude is often expressed as a sum of multiple diversity branches. Thus, it is hard to obtain the closed-form expression of the PDF of $H$. 

Here, we seek for an asymptotically accurate approximation. According to \eqref{out_1} and \eqref{order_define} the asymptotic behaviour of the outage probability depends on $P_{|H|}(x)$ near the origin (when $x \to 0$). Suppose if the diversity order of the overall channel is $d$, then the PDF of $H$ should have a Maclaurin series of:
\begin{equation}\label{mac_should}
P_H(x)=c' \frac{x^{2d-1}}{\gamma_t^d}+o(\frac{x^{2d-1}}{\gamma_t^d}).
\end{equation}
If we take a Laplace transform of this PDF with transform variable $t$, we define:
\begin{align}\label{lap_H_0}
\begin{split}
L_H=\mathcal{L}\{P_{|H|}(x)\}(t)&=\int_{x=0}^\infty e^{-tx}P_{|H|}(x)dx\\
&=\frac{c'\Gamma(2d)}{\gamma_t^dt^{2d}}+o(\frac{1}{\gamma_t^dt^{2d}}).
\end{split}
\end{align}
This result shows that if we expand the Laplace transform of $P_{|H|}(x)$ at $t\to \infty$, the diversity order $d$ can be read off from the first term in the Taylor series. 

\subsection{Perfect phase alignment, Rayleigh links}
With perfect CSI at the RIS and continuous phase shifts, the RIS can be configured to perfectly alignment the phase of the signal from each diversity branch. The resultant magnitude of the effective channel is:
\begin{equation}\label{mag_eff_ch}
|H|=\sum_{m=1}^M |h_m|.
\end{equation}
The next theorem gives a closed-form asymptotic expression for the outage probability.

\begin{theorem}\label{theorem_perfect}
The asymptotic behaviour of the outage probability for perfect phase alignment category can be expressed as:
\begin{equation}\label{asymp_1}
P_{out}(\gamma_0)=\frac{b^{-M}\gamma_0^{M}}{(2M)!}\gamma_t^{-M},
\end{equation}
where $b$ is a constant related to the scale factor of the Rayleigh distributed links.
\begin{proof}
See Appendix~\ref{ap_B}.
\end{proof}
\end{theorem}


According to \eqref{order_define} and \eqref{asymp_1}, we can obtain the diversity order as:
\begin{equation}\label{d1_M}
d = M.
\end{equation}
Additionally, \eqref{asymp_1} can be rewritten as:
\begin{equation}\label{asymp_log}
\log P_{out}=-M\log \gamma_t+ p_0,
\end{equation}
where $p_0=M(\log \gamma_0-\log b)-\log((2M)!)$. This indicate that in high SNR region, the outage probability, plotted on a logarithmic scale, tends to a straight line with a slope of $-M$.

\subsection{Perfect phase alignment, Rician links}
With perfect CSI (at the RIS) and continuous phase shifts, \eqref{mag_eff_ch} still holds in this scenario. Suppose all $|h_m|$ are i.i.d. random variables with Rician distributions:
\begin{equation}\label{h_rician}
P_{|h|}(x)=\frac{x}{b}e^{-\frac{x^2+s^2}{2b}}I_0(\frac{xs}{b}),
\end{equation}
where $s^2$ is the scattered power, $2b$ is the specular power, and $I_0(x)$ is the modified Bessel function of the first kind. The well known shape factor is the ratio of the two: $K=s^2/2b$. The Taylor expansion for this PDF near the origin reads:
\begin{equation}\label{taylor_rice}
P_{|h|}(x)= \frac{e^{-s^2/(2b)}}{b}x + o(x^3).
\end{equation}
We can obtain the Taylor series of $L_H$ at $t \to \infty$:
\begin{equation}\label{taylor_rice_LH}
L_H(t)=\frac{e^{-\frac{Ms^2}{2b}}}{(2b)^M}t^{-2M}+ o(t^{-2M}).
\end{equation}
Using the result in \eqref{lap_H_0}, we obtain the diversity order in this scenario is also $M$.

\subsection{Random phase alignment}
Consider the scenario where CSI is not known at the RIS and random phase shifts are employed. As a result $\theta_m$ is uniformly distribution ed within $[0,2\pi]$. According to \eqref{superposition}, we define $T_c=\sum_{m=1}^M|h_m|\cos\theta_m$, $T_s=\sum_{m=1}^M|h_m|\sin\theta_m$ and investigate their distributions. When the number of diversity branches ($M$) is sufficiently large, according to the central limit theorem, the distributions of $T_c$ and $T_s$ can be approximated by Gaussian distributions with zero means. Moreover, because of the randomness of $\theta_m$, there is no correlation between $T_c$ and $T_s$. As a result, the magnitude of $H$ follow a Rayleigh distribution with a scale factor of $\Omega_p=M$:
\begin{equation}\label{dis_H_ramdom}
P_H(x)=\frac{2x}{\Omega_p}e^{-\frac{x^2}{\Omega_p}}.
\end{equation}
Substituting \eqref{dis_H_ramdom} into \eqref{out_1}, we have in this case:
\begin{align}\label{out_3}
\begin{split}
P_{out}(\gamma_0)&=\int_0^{\sqrt{\gamma_0/\gamma_t}}\frac{2x}{\Omega_p}e^{-\frac{x^2}{\Omega_p}} dx\\
&= 1-e^{-\frac{\gamma_0}{\Omega_p\gamma_t}}.
\end{split}
\end{align}
Then according to \eqref{order_define}, we have:
\begin{equation}\label{d=1}
    d=1.
\end{equation}
Indicating a diversity order of $1$ for the random phase alignment scenario.
\begin{remark}
Although the increasing number of elements (diversity branches) does not improve the diversity order of the combined channel, the average received power does increase with $M$. This fact is reflected in the increase of the scale factor of the Rayleigh distributed overall channel.
\end{remark}

\subsection{Coherent phase alignment}
According to Table~\ref{table2}, various scenarios fall into the same category as the coherent phase alignment. Here, we focus on two typical scenarios: coherent phase alignment due to discrete phase shift RIS and due to the receiver located at positions other than the target direction. The closed-form expression of outage probability in these two scenarios is hard to obtain. Instead of spending the effort to approximate the distribution of the outage probability, we take on another strategy by studying the achievable diversity order of the coherent phase alignment category. As defined in Section~\ref{math_4}, the phase of $h_m$ in coherent phase alignment can be expressed as $\theta_m = \theta_0 + \Delta\theta_m$.
\begin{theorem}\label{theorem_1}
In the coherent phase alignment category, if the following condition applies:
\begin{equation}\label{condition_coherent}
    |\Delta\theta_m-\Delta\theta_n| \leq \pi/2, \ \forall m,n\in [1,M].
\end{equation}
Then, the outage probability can be upper bounded as:
\begin{equation}\label{bound}
P_{out}(\gamma_0)=Pr\{|H|<\sqrt{\frac{\gamma_0}{\gamma_t}} \} < (Pr\{|h_m|<\sqrt{\frac{\gamma_0}{\gamma_t}} \})^M.
\end{equation}
Indicating a full diversity order of $M$ is achieved.
\begin{proof}
The quantity $|\Delta\theta_m-\Delta\theta_n|$ denote the angle between two different branches $\tilde{h_m}$ and $\tilde{h_n}$ on the complex plane. When condition given in \eqref{condition_coherent} holds, this angle could not exceed $\pi/2$. As a result, we have $|H|>|h_m|$ for $\forall m=1,2,\cdots,M$. Consider the $M$-dimensional probability space formed by random variables $h_1, h_2, \cdots h_M$, the volume of $\sum_{m=1}^M|h_m|<\tau$ is smaller than that of $(|h_1|<\tau)\cap(|h_2|<\tau)\cap\cdots\cap(|h_M|<\tau)$. Thus, we proved that \eqref{bound} holds.
\end{proof}
\end{theorem}
With the help of this theorem, we are able to arrive at important results for the following two corollaries.

\begin{corollary}\label{coro_1}
For the discrete phase shift RIS, the achievable diversity order increases with the quantization level $L$. When quantization level $L \geq 4$, full diversity order is achieved.
\begin{proof}
When $L \geq 4$, the angle between two different branches $\tilde{h_m}$ and $\tilde{h_n}$ could not exceed $2\cdot \pi/L=\pi/2$. According to \textbf{Theorem~\ref{theorem_1}}, for Rayleigh or Rician distributed $|h_m|$, the upper bound in \eqref{bound} has a diversity order of $d=M$. This means full diversity order can be achieved.
\end{proof}
\end{corollary}
\begin{remark}
In conclusion, the discrete phase shift RIS has a degraded performance compared with the continuous phase shift RIS, this degradation quickly vanishes as the quantization level increases.
\end{remark}

\begin{corollary}\label{coro_2}
For RIS operating under the 1-D phase scanning configuration, The full-diversity beamwidth (in radians) of the 1-D phase scanning RIS with $M$ elements is approximately $0.25\lambda_c/(Md_x)$, where $\lambda_c$ is the carrier frequency of the signal and $d_x$ is the width of each element.
\begin{proof}
For the 1-D phase scanning, the RIS phase shift is configured according to \eqref{c_scan}. Suppose the beamwidth is $\theta_d$, according to \textbf{Theorem~\ref{theorem_1}}, the maximum phase difference between the diversity branches should satisfy:
\begin{equation}\label{max_dif}
    \Delta\theta = M\cdot 2\pi \sin\theta_d \frac{d_x}{\lambda_c} \leq \frac{\pi}{2}.
\end{equation}
For large $M$, this angle is approximately $\theta_d = 0.25\lambda_c/(Md_x)$

\end{proof}
\end{corollary}

\section{Numerical Results}\label{num}

In this section, numerical results are presented to facilitate the performance evaluations of the RIS-assisted wireless network. We aim to confirm the effectiveness of the proposed analytical results by comparing them with Mount Carlo simulations.

\subsection{Outage Probability for Perfect Phase Alignment Category}
\begin{figure}[ht!]
    \begin{center}
        \includegraphics[scale=0.6]{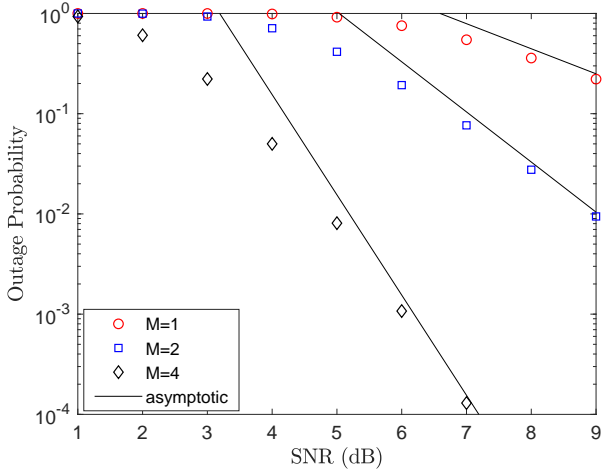}
        \caption{Simulated outage probability and upper bound for Rayleigh branches (from top to bottom: $M$= 1, 2 and 4)}
        \label{out1}
    \end{center}
\end{figure}

Fig.~\ref{out1} plots the outage probability for perfect phase alignment versus the transmit SNR (per branch, in dB). The solid lines represents the asymptotic expression for $M$= 1, 2 and 4, as derived in \eqref{asymp_log}. It can be confirmed that the Mount Carlo simulated points agree with the asymptotic limit in the high SNR region. Moreover, Fig.~\ref{out1} also confirmed the achievable diversity order, which is the negative slope of in the figure, is close to $M$, as derived in \eqref{d1_M}.

\subsection{Outage Probability for Random Phase Alignment Category}
\begin{figure}[ht!]
    \begin{center}
        \includegraphics[scale=0.6]{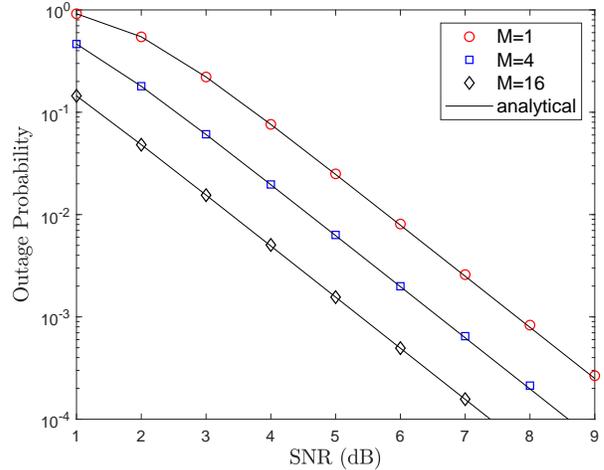}
        \caption{Simulated outage probability for random phase alignment (from top to bottom: $M$= 1, 4 and 16)}
        \label{out2}
    \end{center}
\end{figure}

Fig.~\ref{out2} plots the outage probability for random phase alignment versus the transmit SNR (per branch, in dB). The solid lines represent the analytical expression for $M$= 1, 4, and 16, as derived in \eqref{out_3}. It can be observed that the analytical results fit well with the Mount Carlo simulated points. Fig.~\ref{out2} also confirmed that the achievable diversity order for the random phase alignment case is fixed to one, as derived in \eqref{d=1}. Another observation is that power saving can be achieved from the diversity gain by increasing the number of elements of the RIS. However, the increasing speed of this gain starts to diminish as the number of elements becomes larger. According to Fig.~\ref{out2}, going from $M=1$ to $M=4$, at $1\%$ outage probability there is an approximate $1$ dB reduction in the required SNR. However, if starting from $M=4$, achieving the same $1$ dB SNR reduction requires increasing the number of elements to $M=16$.

\subsection{Outage probability for users located at different angular directions}

\begin{figure}[ht!]
    \begin{center}
        \includegraphics[scale=0.6]{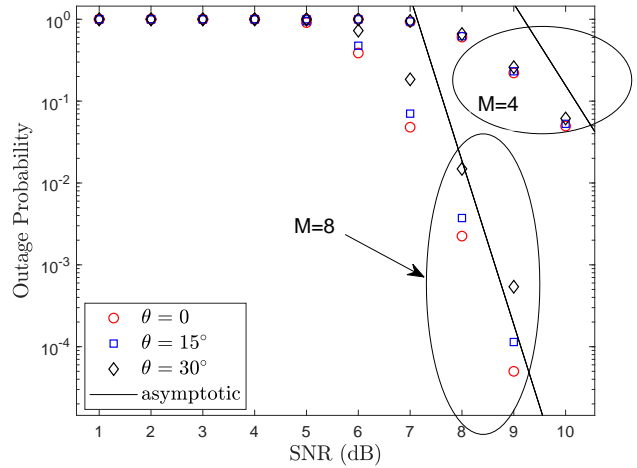}
        \caption{Simulated outage probability for users located at different angular directions.}
        \label{out_ang}
    \end{center}
\end{figure}

Fig.~\ref{out_ang} plot the outage probability for users located at different angular directions w.r.t. the RIS configured base on 1-D phase scanning. The solid lines represent the asymptotic expression of the corresponding perfect phase alignment category for $M=4$ and $8$, as derived in \eqref{asymp_log}. It can be observed that as the direction of the user moves away from the target direction, the outage probability increases. Moreover, this increment is more observable for RIS with a large number of elements ($M$). This is observation is in accordance with \textbf{Corollary~\ref{coro_2}} since the full-diversity order enabling beamwidth decrease with the number of elements as $\Delta\theta \sim M^{-1}$. For example, when $M=8$, this beamwidth is approximately $3.6^\circ$. The Monte Carlo simulated points for $\theta=15^\circ$ and $\theta=30^\circ$ exceed the asymptotic bound as full diversity order can not be achieved outside the beamwidth.

\section{Conclusions}\label{conclusion}

A novel performance analysis was presented from the perspective of phase alignment condition between channels through different RIS elements. Specifically, we proposed four phase alignment categories: perfect alignment, coherent alignment, random alignment, and destructive alignment. With the help of these categories, the performance which users experienced in different directions with respect to RIS can be better analyzed and compared. Under the proposed categorizing approach, performance analysis for the RIS-assisted networks with multiple access schemes remains an open question.

\begin{appendices}
\renewcommand{\theequation}{A.\arabic{equation}}
\setcounter{equation}{0}

      \section{Proof of Expectation Values and Variances for Coherent Phase Alignment Category}\label{ap_A}
      For coherent phase alignment, we consider the scenario where $\theta_m \in (-\pi/L,\pi/L)$. The magnitude of the overall channel ($H$) can be well approximated as Ricianly distributed \cite{xu2020novel}:
      \begin{equation}\label{xu_rician}
          P_{|H|}(x)=\frac{x}{\beta^2}e^{-\frac{x^2+\alpha^2}{2\beta^2}}I_0(\frac{x\alpha}{\beta}),
      \end{equation}
      where the shape factors are related to the quantification level ($L$) of the IRS, the number of elements ($M$) as:
      \begin{align}\label{shape_factors}
          \alpha &= M\bar{h}sinc(\frac{\pi}{2L}),\\
          \beta^2 &= \frac{M}{2}\bar{h^2}[1-sinc(\frac{\pi}{L})].
      \end{align}
      The mean and variance of the distribution can be calculated as:
      \begin{align}\label{m_v}
          \mathbb{E}[H]&=\sqrt{\frac{\pi}{2}}\beta L_{1/2}(-\frac{\alpha^2}{2\beta^2}),\\
          Var[H]&=\alpha^2+2\beta^2-\frac{\pi}{2}\beta^2[L_{1/2}(-\frac{\alpha^2}{2\beta^2})]^2.
      \end{align}
      The proof for other phase alignment categories are straightforward, and are omitted.

      \renewcommand{\theequation}{B.\arabic{equation}}
\setcounter{equation}{0}

      \section{Proof of \textbf{Theorem \ref{theorem_perfect}}}\label{ap_B}
      The critical step is to analysis the behaviour of $\sum_{m=1}^M|h_m|$, which is the sum of M i.i.d. Rayleigh distributed random variables. Here, with the help of Laplace transform and its convolution theorem, the PDF of the Laplace transform of $H=\sum^M |h_m|$ can be derived. First, we denote the Laplace transform of PDFs as follows:
\begin{equation}\label{lap_h}
\mathcal{L}\{P_{|h|}(x)\}(t)=L_h=\int_{x=0}^\infty e^{-tx}P_{|h|}(x)dx,
\end{equation}
\begin{equation}\label{lap_H}
\mathcal{L}\{P_{|H|}(x)\}(t)=L_H=\int_{x=0}^\infty e^{-tx}P_{|H|}(x)dx.
\end{equation}
It is known that the Laplace transform of a convolution is the multiplication of the Laplace transform for each function in the convolution:
\begin{equation}\label{con_theorem}
\mathcal{L}\{f_1*f_2*...*f_M\}=F_1(t)F_2(t)\cdots F_M(t).
\end{equation}
As a result, we have:
\begin{equation}\label{Hh}
L_H(t)=(L_h(t))^M.
\end{equation}
Since $|h|$ follows the Rayleigh distribution:
\begin{equation}\label{rayleigh}
P_{|h|}(x)=\frac{x}{b}e^{-\frac{x^2}{2b}},
\end{equation}
where $b$ is a constant related to the scale factor of the distribution. We can explicitly calculate $L_h$:
\begin{equation}\label{L_h}
L_h = 1-\sqrt{\pi b/2}\cdot te^{bt^2/2}\text{erfc}(t\sqrt{b/2}).
\end{equation}
Base on \eqref{Hh}, we need to preform the inverse Laplace transform to obtain the PDF of $|H|$. Although the PDF for sum of i.i.d. Rayleigh variables does not exist in closed-form, it is sufficient to know only the first few terms in Maclaurin series of $P_{|H|}(x)$ for computations involving small numerical values of $x$. In the light of this, we consider the Taylor expansion of $L_H$ at infinity:
\begin{equation}\label{taylor_inf}
L_H(t)=\sum_{n=1}^{\infty}c_nt^{-n},
\end{equation}
where the index starts at $n=1$. This is the result of $L_H$ being the Laplace transform of $P_H(x)$, thus $L_H \to 0$ when $Re[t] \to +\infty$. $P_{|H|}(x)$ can thus be obtained by performing the inverse transform term by term in the series:
\begin{equation}\label{inv_term}
P_{|H|}(x)= \mathcal{L}^{-1}\{L_H\}=\sum_{n=0}^{\infty}\frac{c_{n+1}}{n!}x^n.
\end{equation}
Using the closed-form expression of $L_h$ in \eqref{L_h} and the relation in \eqref{Hh}, we can obtain the first few term in the Taylor series of $L_H(t)$ as:
\begin{equation}\label{term}
L_H(t)=c_{2M}\cdot t^{-2M}+c_{4M}\cdot t^{-4M}+\cdots,
\end{equation}
where $c_{2M}=b^{-M}$ and $c_{4M}=-3b^{-2M}$. Thus, according to \eqref{inv_term}, the first term in Maclaurin series of $P_{|H|}(x)$ is: 
\begin{equation}\label{term_2}
P_{|H|}(x)=\frac{c_{2M}}{(2M-1)!}x^{2M-1}+O(x^{4M-1}).
\end{equation}
By substituting \eqref{term_2} into \eqref{out_1}, we have:
\begin{equation}\label{out_2}
P_{out}(\gamma_0)=\int_0^{\sqrt{\gamma_0/\gamma_t}}\frac{c_{2M}}{(2M-1)!}x^{2M-1} dx +\cdots.
\end{equation}
Integrating \eqref{out_2} gives us the asymptotic behaviour of the outage probability can be derived.

\end{appendices}
\bibliographystyle{IEEEtran}
\bibliography{mybib}

\begin{thebibliography}{1}
\providecommand{\url}[1]{#1}
\csname url@samestyle\endcsname
\providecommand{\newblock}{\relax}
\providecommand{\bibinfo}[2]{#2}
\providecommand{\BIBentrySTDinterwordspacing}{\spaceskip=0pt\relax}
\providecommand{\BIBentryALTinterwordstretchfactor}{4}
\providecommand{\BIBentryALTinterwordspacing}{\spaceskip=\fontdimen2\font plus
\BIBentryALTinterwordstretchfactor\fontdimen3\font minus
  \fontdimen4\font\relax}
\providecommand{\BIBforeignlanguage}[2]{{%
\expandafter\ifx\csname l@#1\endcsname\relax
\typeout{** WARNING: IEEEtran.bst: No hyphenation pattern has been}%
\typeout{** loaded for the language `#1'. Using the pattern for}%
\typeout{** the default language instead.}%
\else
\language=\csname l@#1\endcsname
\fi
#2}}
\providecommand{\BIBdecl}{\relax}
\BIBdecl

\bibitem{liu2020reconfigurable}
Y.~Liu, X.~Liu, X.~Mu, T.~Hou, J.~Xu, Z.~Qin, M.~Di~Renzo, and N.~Al-Dhahir,
  ``Reconfigurable intelligent surfaces: Principles and opportunities,''
  \emph{arXiv preprint arXiv:2007.03435}, 2020.

\bibitem{xu2020reconfigurable}
P.~Xu, G.~Chen, Z.~Yang, and M.~Di~Renzo, ``Reconfigurable intelligent surfaces
  assisted communications with discrete phase shifts: How many quantization
  levels are required to achieve full diversity?'' \emph{arXiv preprint
  arXiv:2008.05317}, 2020.

\bibitem{wang2020study}
T.~Wang, G.~Chen, J.~P. Coon, and M.-A. Badiu, ``Study of intelligent
  reflective surface assisted communications with one-bit phase adjustments,''
  \emph{arXiv preprint arXiv:2008.09770}, 2020.

\bibitem{wang2020chernoff}
------, ``Chernoff bounds and saddlepoint approximations for the outage
  probability in intelligent reflecting surface assisted communication
  systems,'' \emph{arXiv preprint arXiv:2008.05447}, 2020.

\bibitem{ding2020impact}
Z.~Ding, R.~Schober, and H.~V. Poor, ``On the impact of phase shifting designs
  on irs-noma,'' \emph{arXiv preprint arXiv:2001.10909}, 2020.

\bibitem{mailloux2017phased}
R.~J. Mailloux, \emph{Phased array antenna handbook}.\hskip 1em plus 0.5em
  minus 0.4em\relax Artech house, 2017.

\bibitem{xu2020novel}
J.~Xu and Y.~Liu, ``A novel physics-based channel model for reconfigurable
  intelligent surface-assisted multi-user communication systems,'' \emph{arXiv
  preprint arXiv:2008.00619}, 2020.

\end{thebibliography}

\end{document}